\documentclass[twocolumn]{aastex631}
\usepackage{color}
\usepackage{graphicx}
\usepackage{amsmath}
\usepackage{threeparttable}
\usepackage{bm}

\newcommand{\Mpc}{\mathrm{~km~s^{-1}~Mpc^{-1}}}

\begin{document}
\title{Newest measurements of Hubble constant from DESI 2024 BAO observations}

\author{Wuzheng Guo}
\affiliation{Institute for Frontiers in Astronomy and Astrophysics, Beijing Normal University, Beijing 102206, China;}
\affiliation{School of Physics and Astronomy, Beijing Normal University, Beijing 100875, China;}

\author{Qiumin Wang}
\affiliation{Institute for Frontiers in Astronomy and Astrophysics, Beijing Normal University, Beijing 102206, China;}
\affiliation{School of Physics and Astronomy, Beijing Normal University, Beijing 100875, China;}

\author{Shuo Cao$^{\ast}$}
\affiliation{Institute for Frontiers in Astronomy and Astrophysics, Beijing Normal University, Beijing 102206, China;}
\affiliation{School of Physics and Astronomy, Beijing Normal University, Beijing 100875, China;}
\email{$\ast$ caoshuo@bnu.edu.cn}

\author{Marek Biesiada$^{\dagger}$}
\affiliation{National Centre for Nuclear Research, Pasteura 7, PL-02-093 Warsaw, Poland;}
\email{$\dagger$ Marek.Biesiada@ncbj.gov.pl}

\author{Tonghua Liu$^{\ddagger}$}
\affiliation{School of Physics and Optoelectronic, Yangtze University, Jingzhou 434023, China;}
\email{$\ddagger$ liutongh@yangtzeu.edu.cn}

\author{Yujie Lian}
\affiliation{Institute for Frontiers in Astronomy and Astrophysics, Beijing Normal University, Beijing 102206, China;}
\affiliation{School of Physics and Astronomy, Beijing Normal University, Beijing 100875, China;}

\author{Xinyue Jiang}
\affiliation{Institute for Frontiers in Astronomy and Astrophysics, Beijing Normal University, Beijing 102206, China;}
\affiliation{School of Physics and Astronomy, Beijing Normal University, Beijing 100875, China;}

\author{Chengsheng Mu}
\affiliation{Institute for Frontiers in Astronomy and Astrophysics, Beijing Normal University, Beijing 102206, China;}
\affiliation{School of Physics and Astronomy, Beijing Normal University, Beijing 100875, China;}

\author{Dadian Cheng}
\affiliation{Institute for Frontiers in Astronomy and Astrophysics, Beijing Normal University, Beijing 102206, China;}
\affiliation{School of Physics and Astronomy, Beijing Normal University, Beijing 100875, China;}

\begin{abstract}


In this Letter, we use the latest results from the Dark Energy Spectroscopic Instrument (DESI) survey to measure the Hubble constant. Baryon acoustic oscillation (BAO) observations released by the DESI survey, allow us to determine $H_0$ from the first principles. Our method is purely data-driven and relies on unanchored luminosity distances reconstructed from SN Ia data and $H(z)$ reconstruction from cosmic chronometers. Thus it circumvents calibrations related to the value of the sound horizon size at the baryon drag epoch or intrinsic luminosity of SN Ia. We find $H_0=68.4^{+1.0}_{-0.8}~\Mpc$ at 68\% C.L., which provides the Hubble constant at an accuracy of 1.3\% with minimal assumptions. Our assessments of this fundamental cosmological quantity using the BAO data spanning the redshift range $z=0.51-2.33$ agree very well with Planck's results and TRGB results within $1\sigma$. This result is still in a $4.3\sigma$ tension with the results of the Supernova H0 for the Equation of State (SH0ES).

\end{abstract}


\section{Introduction}

As one of the most fundamental cosmological parameters, the Hubble constant ($H_0$) plays an important role in understanding our universe, especially its current expansion rate, composition, and ultimate fate. However, the $H_0$ values deduced from observations of the early and late Universe do not agree with each other. In the last decade, this so-called "Hubble tension" has become an intriguing problem in modern cosmology. Type Ia Supernovae (SN Ia) and cosmic microwave background radiation (CMB) are two well-established cosmological probes. The SN Ia based measurement follows from the astronomical distance ladder in the local universe, while CMB, which froze the temperature fluctuation after the Big Bang, infers $H_0$ value from the inverse distance ladder based on a cosmological model, being $\Lambda$CDM the assumed standard model. Specifically, the SH0ES project favored a higher value of $H_0 = 73.04\pm1.04~\Mpc$ \citep{SH0ES2022}, while the Planck collaboration (CMB measurements) provided a different lower value of $H_0=67.4\pm0.5~\Mpc$ \citet{Planck2020}. Currently, the Hubble tension has reached a significance level of $5-6\sigma$, making it difficult to explain as a mere statistical fluke. Accordingly, there is intense work in the literature discussing whether our current standard cosmological model ($\Lambda$CDM) needs to be replaced with the new physics, e.g., interacting dark energy (IDE) models \citep{ Farrar2014, Valentino2017, Yang2018}, f(T) theories \citep{Starobinsky2007, Bengochea2011}, etc. In order to alleviate the $H_0$ tension, great efforts have also been made to identify and overcome systematic effects that can affect the astrophysical distance measurements \citep{Rigault2015, Spergal2015, Vagnozzi2020, LiuYang2023}. In this respect, the emergence of completely new cosmological-model-independent methods for inferring the value of $H_0$ are very important. For instance, \citet{Bernal2016} reconstructed the late time expansion history and extrapolated towards $H_0$ using Baryon Acoustic Oscillations (BAO) and SN Ia data. It should be mentioned that their reconstructions rely on the sound horizon scale $r_{\rm d}$ at the radiation drag epoch, which would be an extra parameter brought into the analysis. Recently, \citet{Renzi2023} proposed a new methodology to constrain $H_0$ only based on the distance duality relation (DDR hereafter) and direct standard observations, i.e. SN Ia, BAO, and cosmic chronometers (CC) data. In this paper, inspired by the latest Dark Energy Spectroscopic Instrument (DESI) survey data release, we follow their work and update the constraints on the Hubble constant at different redshifts, trying to find new clues to fix the Hubble tension.

Around the recombination epoch of our universe, when the photons decoupled from baryonic particles, the features of matter clustering were imprinted on the matter distribution and stretched with the expansion of the universe. Nowadays, we can observe these features by measuring the galaxy correlation functions and finding a single localized peak in it. The scale corresponding to the peak position is the comoving galaxy separation of $r_{\rm d} \sim 150~{\rm Mpc}$, which makes BAO a standard ruler in cosmic research. Apparently, the measurement of the accurate value of this scale relied on a large field sky survey project, so Sloan Digital Sky Survey (SDSS) gave us the first BAO feature measurement in 2005 \citep{SDSS2005}. A few years later, a sample containing $11$ data from BAO observations was established \citep{Beutler2011, Font2014, Ross2015, Alam2017, Bautista2017, Ata2018, Ryan2019}. This sample initiated great progress in constraining cosmological models \citep{Lian2021, Guo2022}. However, not all the data in that sample could be used in our research since some of their measurements could not get rid of the effect of $r_{\rm d}$, which would introduce unnecessary priors.

As a successor of the previous SDSS survey, Data Release 1 (DR1) from the new DESI survey, which aimed at scanning over $14200$ squared degrees in the redshift range $0.1<z<4.2$ for five years of observations, has produced exciting results from BAO measurements. A recent DESI data release provided five new BAO measurements with transverse co-moving distance $D_{\rm M}$ and co-moving distance along the line of sight $D_{\rm H}$ which made it possible to eliminate $r_{\rm d}$ \citep{DESI2024, DESI-Lalpha, DESI-QSO&Gal}. We use this opportunity in our constraints, since $r_{\rm d}$ derived from CMB observations would bias our results. The DESI data relevant to our work was measured from luminous red galaxies (LRG), emission-line galaxies (ELG) as direct tracers, and indirect Lyman-$\alpha$ (Ly$\alpha$) forest quasars tracing through the distribution of neutral hydrogen. Combined with the other two tracers from the low redshift galaxies of bright galaxy sample (BGS) and QSO, which we did not use in our analyses, \citet{DESI2024} required $H_0=68.52\pm0.62\Mpc$ with priors from Big Bang nucleosynthesis and CMB. This new data release also promoted testing new ideas for relieving the Hubble tension \citep{Dlifton2024}.

\section{Method}\label{Sec:2}

The Hubble constant is directly linked to the distances measured from extragalactic objects within our universe. More precisely, these distances can be further categorized into luminosity distance $d_{\rm L}$ and angular diameter distance $d_{\rm A}$. The former is derived from the observed emission flux of a source of known intrinsic luminosity, while the latter is based on knowledge of the intrinsic angular size of the object. These two types of distances are linked together through the DDR, which is expressed as
\begin{equation}
    \frac{d_{\rm L}(z)}{(1+z)^2d_{\rm A}(z)}=1.
    \label{Eq.1}
\end{equation}
Such relation is valid in any metric theory of gravity and requires two conditions needed to be met: that photons propagate along null geodesics, and their number is conserved during the propagation. Various tests have been done so far to verify the validity of the DDR by testing the above ratio \citep{2011RAA....11.1199C, 2016MNRAS.457..281C, 2019PhRvD..99f3507Q, Zheng2020,2023PhLB..83837687L} consistently justified the validity of Eq.~(\ref{Eq.1}).

Now the Hubble constant can be formulated as
\begin{equation}
    H_0=\frac{1}{(1+z)^2}\frac{\Xi_{\rm SN Ia}(z)}{\Xi_{\rm BAO}(z)}H_{\rm CC}(z),
    \label{Eq.2}
\end{equation} 
which expresses $H_0$ in terms of independently observable quantities. On the one hand, $\Xi_{\rm SN Ia}(z)=H_0 d_{\rm L}(z)$ is the unanchored luminosity distance, associated with SN Ia observations through their measurements of apparent magnitude. On the other hand, $\Xi_{\rm BAO}(z)=H(z)d_A(z)$ is the product of the Hubble parameter ($H(z)$) and the angular diameter distance ($d_{\rm A}(z)$). Especially, $d_{\rm A}(z)$ can be calculated by combining the transverse and line-of-sight co-moving distances from BAO measurements, without relying on the value of sound horizon $r_{\rm d}$ at the drag epoch. The cosmic chronometers (CC) offer a direct measurement of the Hubble parameter $H(z)$ by using the differential ages of galaxies, which are determined by the evolution of passively aging star populations and their spectroscopic redshifts.

Five data points regarding BAO measured by DESI are central to our implementation of Eq.~(\ref{Eq.2}). Therefore, the other data sets i.e. standard clocks (CC), and standard candles (SN Ia) should provide the necessary ingredients at the redshifts corresponding to the BAO measurements. Simple interpolation would introduce too much uncertainty and bias to the final results. Therefore, in our paper, we use the Gaussian Processes (GP) regression to reconstruct SN Ia and CC data at BAO's redshifts. This approach is different from traditional GP by assuming that observational data adhere to N-dimensional Gaussian distributions centered around an input prior mean function \citep{ Liao2019, Liao2020, Li2024, LiuLiao2024, LiuZhong2024}. 
We choose $\Lambda$CDM model as the mean function in reconstructions (with the fiducial value of matter density parameter $\Omega_{\rm m}=0.30$). Then based on the same prior input mean function, different input hyperparameters in GP regression would reconstruct different curves about the required variable, and the best 1000 curves that match the data, whose $\chi^2$ are all smaller than the mean function, are selected as the final reconstruction result in this paper. 

\begin{figure*}[ht]
    \centering
        \includegraphics[width=0.32\linewidth]{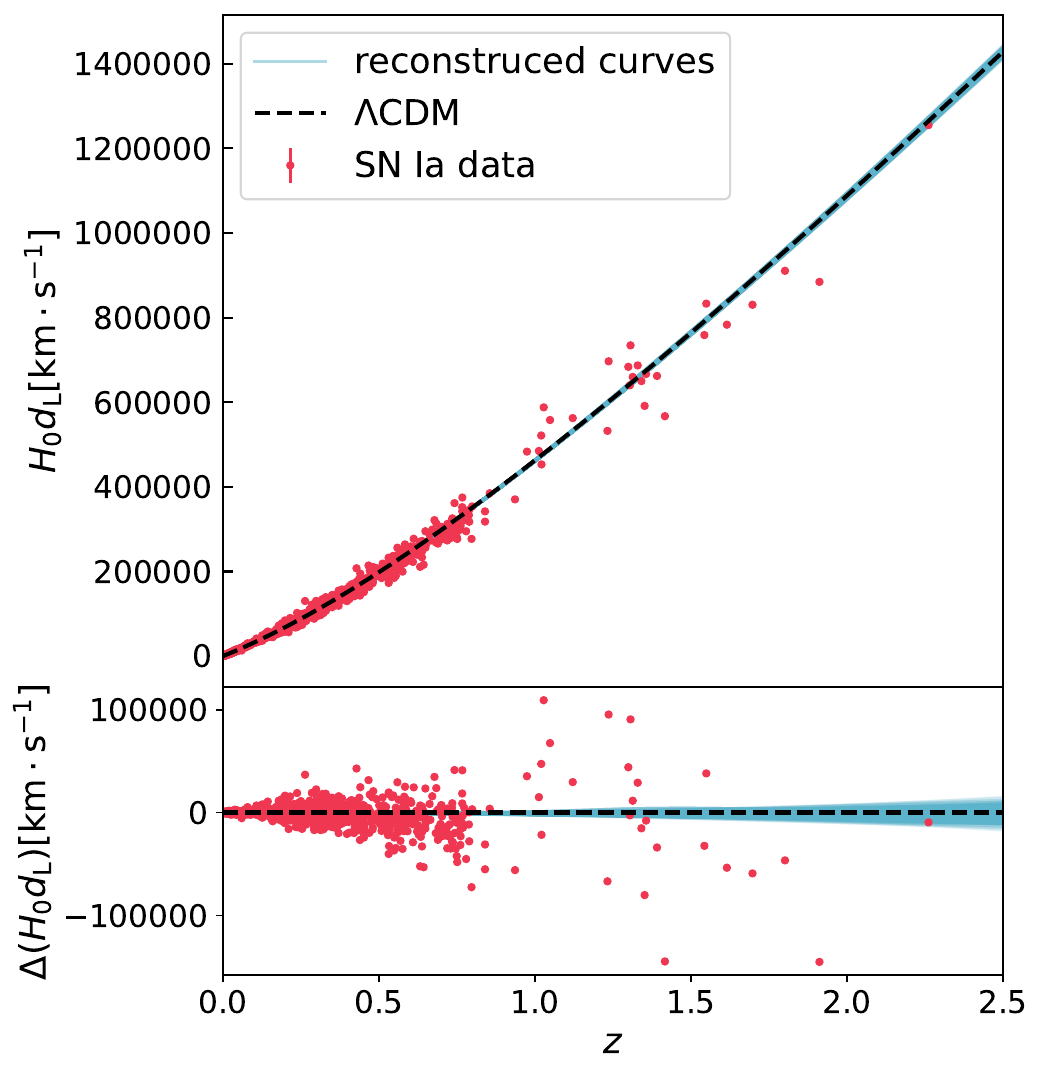}
        \includegraphics[width=0.32\linewidth]{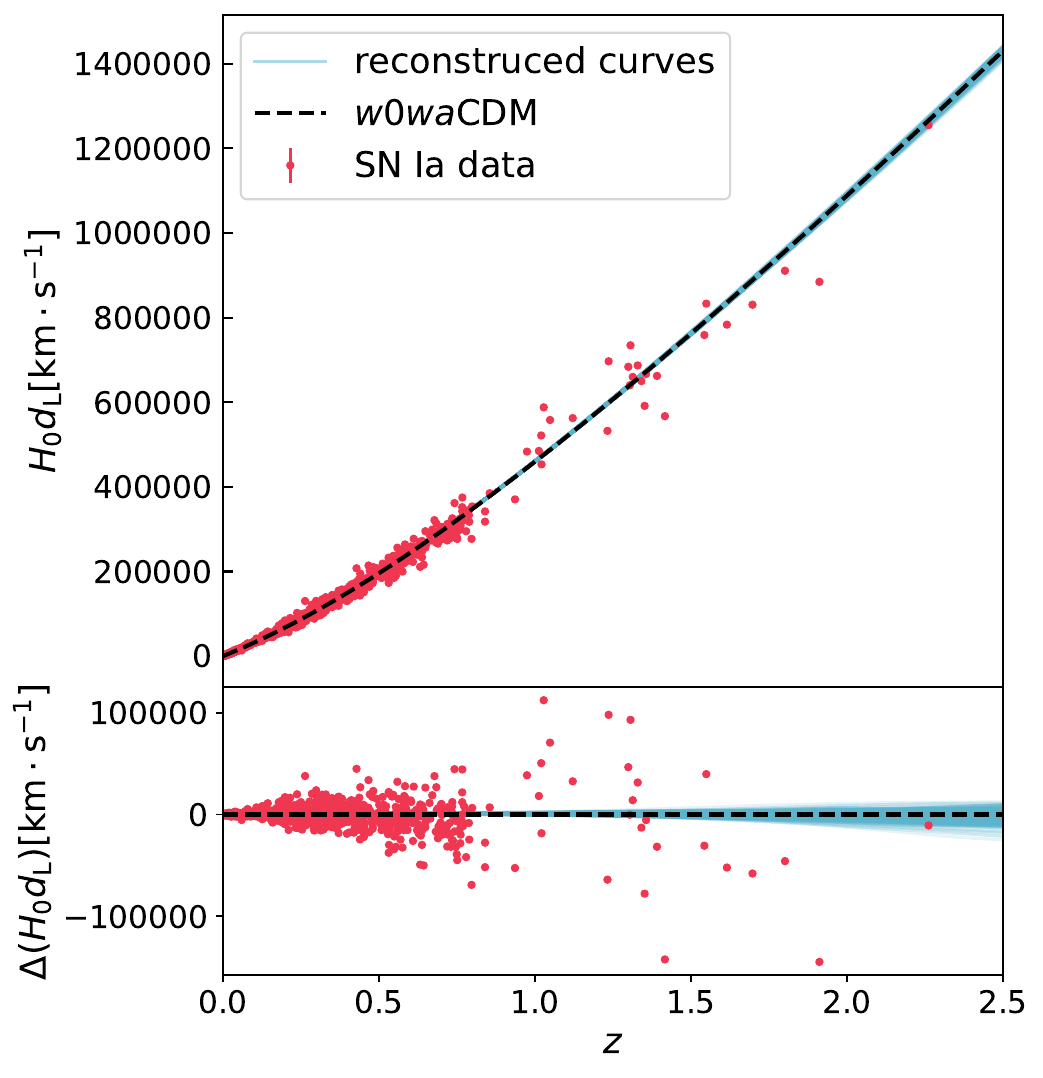}
        \includegraphics[width=0.32\linewidth]{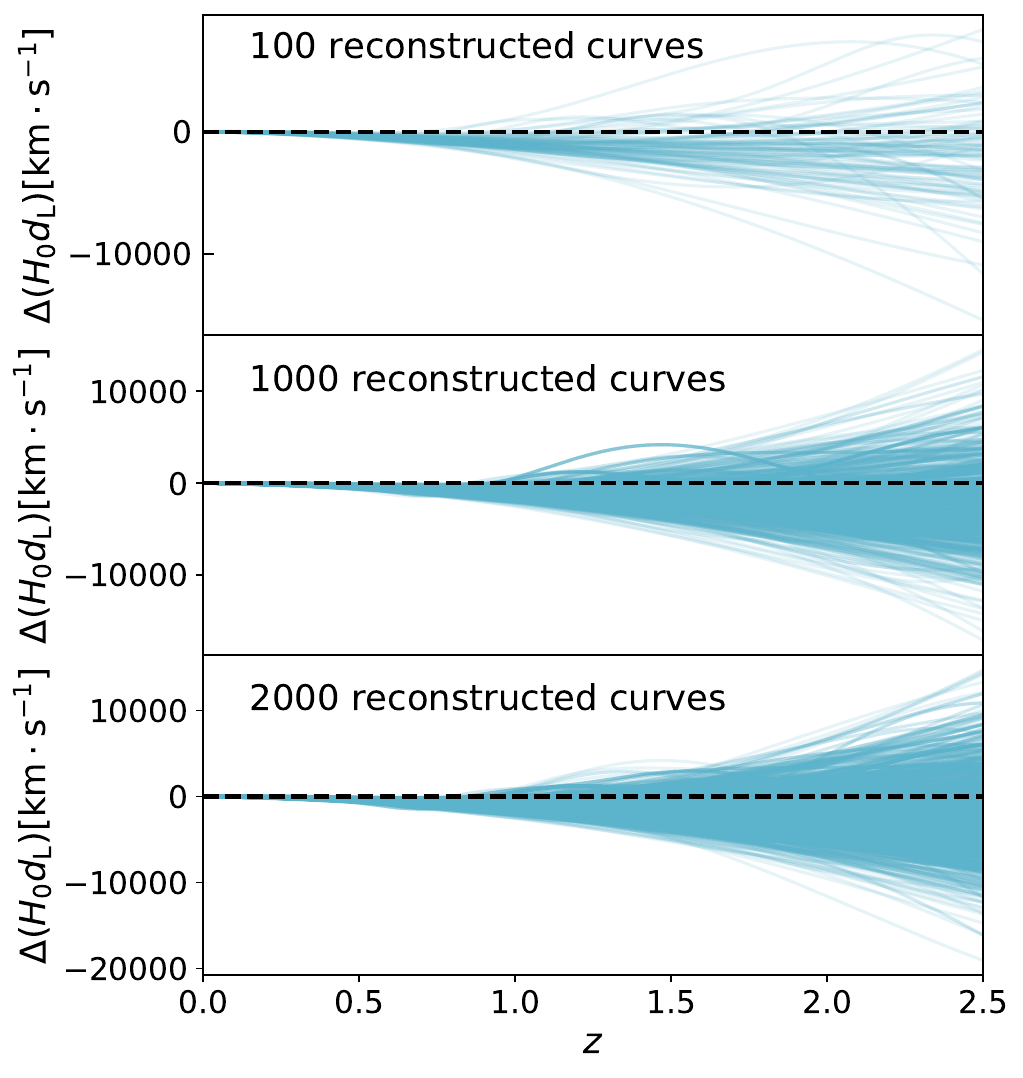}
    \caption{\textbf{Left panel:} Reconstructed $H_0d_{\rm L}$ from SN Ia data using GP regression, with fiducial $\Lambda$CDM model as a prior. \textbf{ Middle panel:} Reconstructed $H_0d_{\rm L}$ from SN Ia data using GP regression, with fiducial $w_0w_a$CDM model as a prior. The residuals between reconstructed $H_0d_{\rm L}$ and their input mean functions are also shown in each bottom panel. \textbf{Right panel:} The residuals between reconstructed $H_0d_{\rm L}$ and the fiducial $\Lambda$CDM with different line numbers. The red dots with the error bar are the SN Ia dataset used in reconstruction, the blue curves show our reconstructed results, and the black dashed line presents the mean function in each reconstruction.}
    \label{fig:1}
\end{figure*}

\section{Data}

Our analysis encompasses three publicly accessible datasets: five BAO data points from the DESI Data Release 1 \citep{DESI2024}; the SN Ia sample from the PantheonPlus dataset \citep{SH0ES2022}; and a compilation of 32 CC measurements of the Hubble parameter $H(z)$ \citep{Qi2023}.

At the drag epoch of the early universe, the photons decoupled from the baryons, thus the primary disturbance propagating at the speed of sound was imprinted on the matter distribution. This feature stretched with the expansion of the universe and appeared as the inhomogeneous distribution of the luminous objects. So in BAO measurement, this distribution can be measured by counting the number of galaxy pairs at different separation scales and distinguishing the scale related to the peak of the two-point correlation function as the BAO scale at that redshift. Measurable quantities comprise the preferred angular separation of galaxies in the direction perpendicular to the line of sight and preferred redshift separation in the direction along the line of sight. Hence the ratios $D_{\rm M}/r_{\rm d}$ and $D_{\rm H}/r_{\rm d}$ are observables and in order to use the distances $D_{\rm M}$ and $D_{\rm H}$ one needs to know the size of the sound horizon $r_{\rm d}$. However, thanks to the DDR, there is no need to use these two distances individually since their ratio can be transformed to $H(z)d_{\rm A}(z)$ which is the same as the denominator of Eq.~(\ref{Eq.2})
\begin{equation}
    \frac{D_{\rm M}}{D_{\rm H}}=\frac{(1+z)d_{\rm A}(z)}{cH(z)^{-1}}=\frac{1+z}{c}H(z)d_{\rm A}(z),
    \label{Eq.3}
\end{equation}
where $c$ is the speed of light. 

Up to now, the latest DR1 of DESI only covered half of the expected area of the survey, while it already provided more than twice the number of redshifts compared with SDSS, achieving a higher precision (about 0.49\%) on the BAO isotropic scale \citep{DESI2024}. Due to the possible limitations of the survey time, the signal-to-noise ratio of BGS and QSO tracers was not high enough so only angle averaged signal was measured. Therefore, only five data points from LRG, ELG, and Ly$\alpha$ tracers are available for our purpose of estimating $H_0$. Due to its high precision, this dataset has already been widely used in studying the properties of dark energy \citep{Cort2024, Giar2024, Wang2024}, trying to release the Hubble tension \citep{Bousis2024, Jia2024}, testing modified gravity theory \citep{Escamilla2024}. 

Regarding Type Ia Supernovae (SN Ia), the Pantheon+ dataset has been significantly expanded in both sample size and redshift coverage compared to its predecessor, the Pantheon dataset \citep{Scolnic2018}. The distance modulus was derived from the Pantheon+SH0ES dataset, which included $1701$ light curve measurements from $1550$ distinct supernovae spanning a redshift range of $z\in[0.001,2.26]$. Additionally, the host distance of Cepheid variable stars was determined from $77$ data points associated with supernovae in their host galaxies, with a redshift range of $0.00122<z<0.01682$. The luminosity distance of SN Ia is related to the distance modulus given by 
\begin{equation}
\mu_{\rm SN} \equiv m_{\rm B} - M_{\rm B} = 5\log_{10}\left(\frac{d_{\rm L}}{{\rm Mpc}}\right) + 25,
\label{Eq.4}
\end{equation}
where $m_{\rm B}$ represents the observed magnitude in the rest-frame B-band and $M_{\rm B}$ is the absolute magnitude. Note that $M_{\rm B}$ is tightly correlated with $H_0$. Therefore, following \citep{Riess_2016} we introduce a calibrating term $a_{\rm B} = \log_{10} H_0 - 0.2 M_{\rm B} -5$ instead of the absolute magnitude $M_{\rm B}$ and obtain
\begin{equation}
    H_0d_{\rm L}(z)=10^{0.2m_{\rm B}+a_{\rm B}}.
    \label{Eq.5}
\end{equation}
This unanchored luminosity distance effectively avoids the bias introduced by the degeneracy between $H_0$ and $M_{\rm B}$, and the calibration term was also constrained at $a_{\rm B} = 0.71273\pm0.00176$ \citep{Riess_2016}. Hence we follow this approach in our reconstruction and the uncertainty of $a_{\rm B}$ is included in order to get a robust conclusion. Thus all the uncertainties in our GP regression reconstruction could be treated using the covariance matrix,  which includes both statistical and systematic uncertainties.

For SN Ia, systematic effects can induce correlations among different supernovae, and thus, the optimal fitting point should be adjusted for their presence. Ignoring these effects not only underestimates the uncertainty but may also bias the results \citep{Conley2011}. Consequently, we utilize the covariance matrix to marginalize all systematic terms during the fitting process. The reconstructed unanchored luminosity distance ($H_0d_{\rm L}$) is depicted as blue curves in the left panel of Fig.~\ref{fig:1}, with the dashed black line representing the $\Lambda$CDM model. We also show the reconstruction results with the $w_0w_a$CDM model acting as the mean function (with the dark energy equation of state parameters derived in \citet{WangHu2024}), since both DESI BAO and Pantheon+ data actually allows for a $w_0w_a$CDM model with $w_0>-1$ and $w_a<0$. The reconstructed curves with different mean functions prefer almost the same reconstruction results, which confirms again that the GP regression we use is essentially a model-independent method in the sense that the mean functions only act as a prior in GP regression. Moreover, the reconstruction results with different numbers of lines ($N=100, 1000, 2000$) are also displayed in Fig.~\ref{fig:1}. Our results suggest that reconstruction curves follow the N-dimensional Gaussian distribution, hence the larger number of curves used for calculating $H_0$ would only help to smooth the posterior distribution and does not improve the resulting mean value of $H_0$ nor its $1\sigma$ range \citep{Li2021, LiuLiao2024}. To sum up, using 1000 reconstruction lines with the fiducial $\Lambda$CDM model as the mean function is suitable for the subsequent calculation. For the purpose of calculating $H_0$, we use reconstructed unanchored luminosity distances corresponding to the redshifts of the BAO data points.

\begin{figure}
    \centering
    \includegraphics[width=0.9\linewidth]{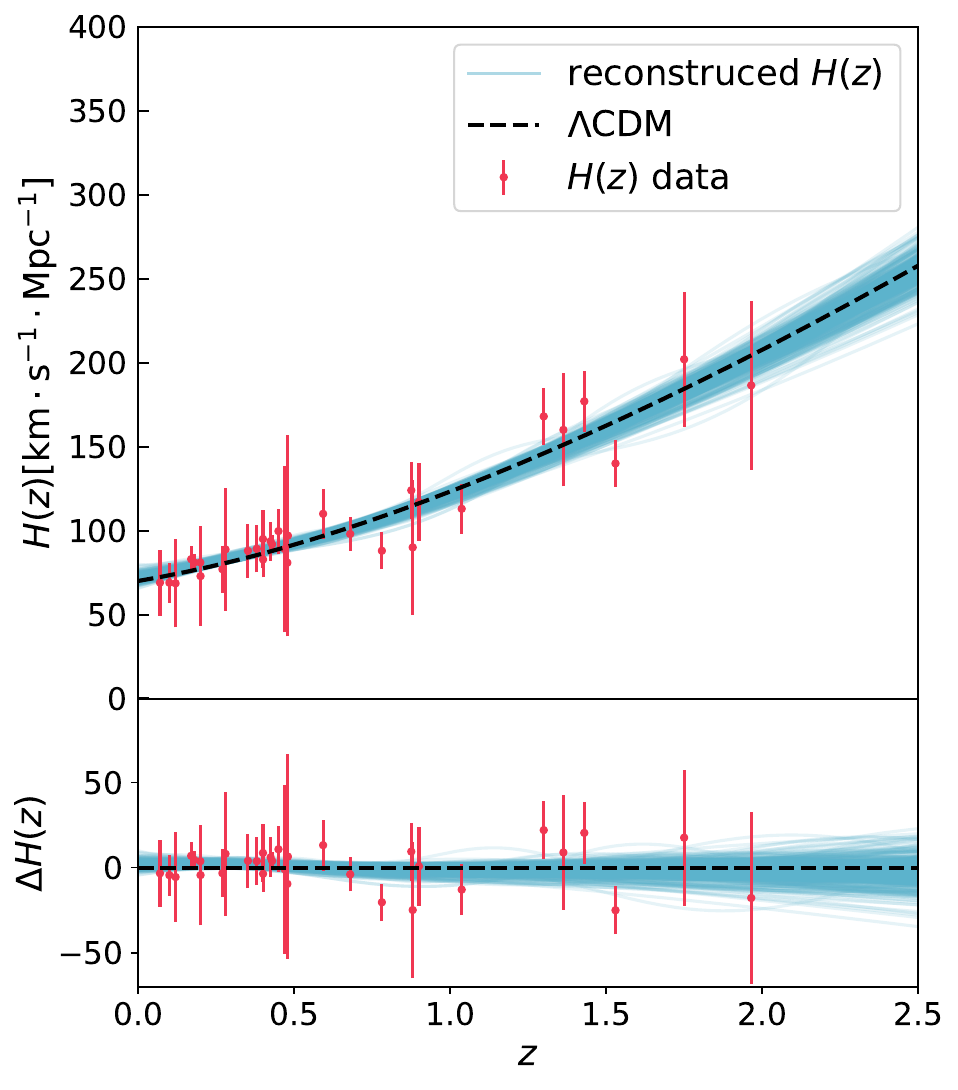}
    \caption{Reconstructed $H(z)$ from CC data using GP regression. The black dashed line shows the fiducial $\Lambda$CDM model. The residuals between reconstructed $H(z)$ and the fiducial $\Lambda$CDM are also shown in the bottom panel.
    }
    \label{fig:2}
\end{figure}

The last ingredient we need is the Hubble parameter $H(z)$, which can be expressed as 
\begin{equation}
    H(z)=-\frac{1}{1+z}\frac{{\rm d}z}{{\rm d}t},
    \label{Eq.6}
\end{equation}
where the redshift $z$ could be obtained from spectroscopic surveys with high accuracy and the differential age evolution of the Universe (${\rm d}t$) was derived from the age evolution of massive galaxies. Different from other galaxies, massive ($M_{\rm stars}>10^{11}M_\odot$) early-type galaxies formed $>90\%$ of their stellar mass rapidly (at high-redshift $z>2-3$), and have no subsequent major episode of star formation. Therefore, this type of galaxy provided an ideal environment to measure $H(z)$ due to its age and stable evolution \citep{Moresco2016}. This method based on observational $H(z)$ data (OHD) has been widely used to test cosmological models \citep{Zhang2014}. 

In this paper, we use the most recent data set, encompassing $32$ cosmic chronometers \citep{Qi2023}, which spans a redshift range from $z=0.07$ to $z=1.965$. This approach holds promise for determining the Hubble constant independently of cosmological models, although it is imperative to meticulously assess the associated system uncertainties \citep{Moresco2020}. Since there were still some uncertainties in the determination of the physical properties of galaxies, both statical errors and systematic errors were considered in this dataset \citep{Qi2023}.

The results of our reconstructions, based on this comprehensive sample, are presented in Fig.~\ref{fig:2}. It is evident that our reconstructed data for both SN Ia and CC align well with the predictions of the $\Lambda$CDM model. This concordance reaffirms our approach to adopt the $\Lambda$CDM model as a prior in the GP regression analysis, which could be considered a prudent choice for our study.

\begin{figure}
    \centering
    \includegraphics[width=1\linewidth]{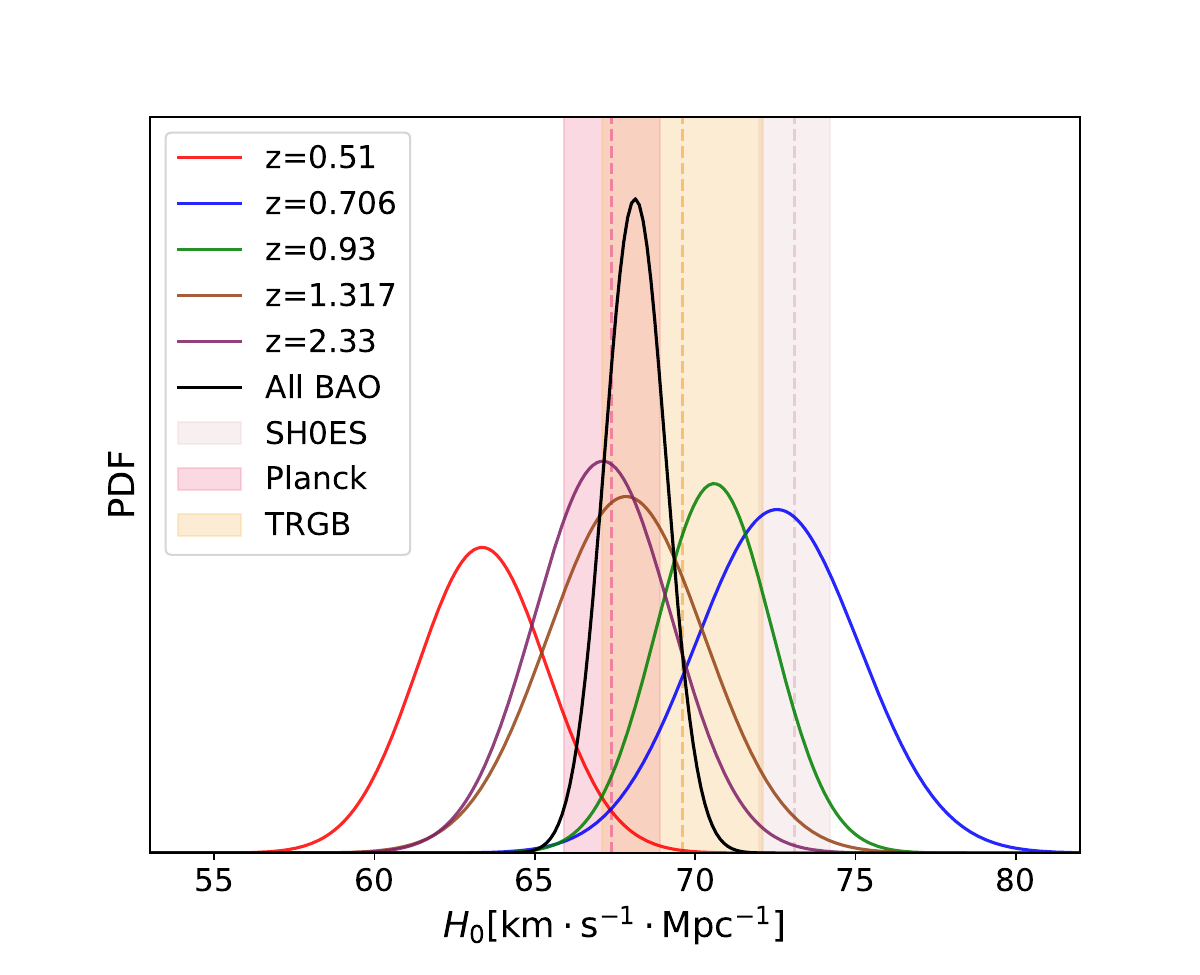}
    \caption{Posterior distribution functions (PDFs) of the Hubble constant $H_0$. Shaded regions represent the constraints on $H_0$ and their $1\sigma$ intervals from SH0ES \citep{SH0ES2022}, Planck \citep{Planck2020} and TRGB \citep{TRGB2020} }
    \label{fig:3}
\end{figure}

\begin{table}
    \renewcommand\arraystretch{1.8}
    \centering
    \caption{Value of the Hubble constant at different redshifts corresponding to BAO data points from DESI. The joint posterior $H_0$ value is denoted as total.}
    \begin{tabular}{cc}
    \hline
        $z$ & $H_0(\Mpc)$\\
    \hline
        total   & $68.4^{+1.0}_{-0.8}$\\
        $0.51$  & $63.4^{+2.3}_{-2.1}$\\
        $0.706$ & $72.4^{+2.8}_{-2.5}$\\
        $0.93$  & $70.8\pm1.8$\\
        $1.317$ & $67.9^{+2.9}_{-2.4}$\\
        $2.33$  & $67.5\pm2.3$\\
    \hline
    \end{tabular}
    
    \label{tab:1}
\end{table}

\section{Results and discussion}\label{Sec:3}

Our approach to constrain the Hubble constant from BAO data supplement by SN Ia unanchored distances and CC can be outlined as follows. \textbf{I)} Reconstruct the unanchored luminosity distance $H_0d_{\rm L}(z)$ relation, and $H(z)$ relation using available compilations of SN Ia data (PantheonPlus sample) and CC, respectively. GP regression technique is used for this purpose. \textbf{II)} Our reconstruction contains 1000 $H_0d_{\rm L}(z)$ and 1000 $H(z)$ curves from which the values corresponding to BAO redshifts are picked. Then, at each BAO redshift the value of $H(z)d_{\rm A}(z)$ is sampled randomly from the Gaussian distribution representing the measured values and their uncertainties given by DESI. This way, $1000\times1000$ possible values for $H_0$ are generated for each redshift. From these values one-dimensional posterior distributions (PDFs) of $H_0$ at five distinct BAO redshifts are derived. \textbf{III)} Multiply all five PDFs together to establish a comprehensive joint constraint on $H_0$. The advantage of this paper is that our reconstructions of necessary ingredients are solely data-driven and we do not need to rely on calibrating the absolute magnitude of SN Ia. 

Incorporating the most recent data from BAO, we determine the Hubble constant as $H_0=68.4^{+1.0}_{-0.8}~\Mpc$ (median value plus the $16^{\rm th}$ and $84^{\rm th}$ percentiles around it). The PDFs of the Hubble constant at five different BAO redshifts and the combined result are depicted in Fig.~\ref{fig:3}, with the numerical constraints detailed in Table \ref{tab:1}. Our estimate of $H_0$ is well consistent with the latest results of $H_0=68.5\pm1.5~\Mpc$, based on a combination of independent geometrical data sets \citep{Renzi2023}. Moreover, the enhanced precision in DESI BAO data has enabled us to achieve a more stringent joint constraint on $H_0$, despite using fewer BAO data points compared with their results. It is noteworthy that the redshift of Lyman-$\alpha$ point in BAO data exceeds the redshift ranges of both SN Ia and CC, which would result in some bias when we extend GP regression results. If the Lyman-$\alpha$ point is not considered in the analysis, the corrected joint estimate is adjusted to $68.6\pm1.0~\Mpc$, which also agrees with the recent findings obtained under the same assumption \citep{Renzi2023}. DESI collaboration pointed out that their BAO measurements regarding the LRG tracer showed a 3$\sigma$ tension with the SDSS measurements in the redshift range $0.6<z<0.8$. The cosmological constraints showed a 2$\sigma$ offset in $F_{\rm AP}$ between the LRG data point in $0.4<z<0.6$ and the $\Lambda$CDM expectation \citep{DESI2024}. Therefore, we present here the effect of them on the constraints of Hubble constant: $H_0=69.7^{+1.1}_{-1.0}~\Mpc$ without LRG1 point, $H_0=67.6\pm1.0~\Mpc$  without LRG2 point and $H_0=69.1^{+1.2}_{-1.1}~\Mpc$ without LRG1 and LRG2 points. It is clear that the inclusion of the LRG tracer could introduce a non-negligible influence on the joint result. Specially, in the future analysis of DESI 3-year and 5-year data, one should pay more attention to the monopole component in LRG2 \citep{WangHu2024}. For the purpose of determining one element necessary for our assessment, i.e., unanchored luminosity distances, we use the Pantheon+ sample of SN Ia. The Pantheon+SH0ES data alone used as cosmological probes \citep{SH0ES2022} preferred much higher values of $H_0 = 73.4\pm 1.1\Mpc$, $73.5\pm 1.1~\Mpc$ and $73.3\pm 1.1~\Mpc$, for the $\Lambda$CDM, flat $w$CDM, and flat $w_0w_a$CDM models, respectively. One might have worried whether this data set could leverage $H_0$ inferred to some higher values.

Our findings indicate that on the contrary, our $H_0$ estimate remains consistent with the early Universe's coherent model \citep{Planck2020} and its expansion history \citep{Eisenstein2005}. Let us remind you that we use BAO alone, without the need to invoke CMB for calibrating the sound horizon scale.  

Our assessments of this fundamental cosmological quantity, based on the BAO data spanning the redshift range of $z=0.51-2.33$, agree very well with both Planck's results and TRGB results within $1\sigma$ \citep{TRGB2020}. However, there is still a $4.3\sigma$ tension between our measurements and the results of SH0ES) \citep{SH0ES2022}. The redshifts probed by DESI are low as compared with CMB measurements, hence the Hubble tension seems to be not so much about low vs. high redshift observations, but the physics of early vs. late Universe.  

\section*{Acknowledgements}

This work is supported by the National Natural Science Foundation of China (Nos. 12021003, 12203009, 12433001); The Beijing Natural Science Foundation No. 1242021, the Strategic Priority Research Program of the Chinese Academy of Sciences, Grant No. XDB23000000; The Interdiscipline Research Funds of Beijing Normal University,  and the Fundamental Research Funds for the Central Universities; The Chutian Scholars Program in Hubei Province (X2023007); Hubei Province Foreign Expert Project (2023DJC040).

\bibliography{reference.bib}{} 
\bibliographystyle{aasjournal}

\end{document}